\documentclass[aip,apl,amsmath,amssymb,reprint,]{revtex4-1}
\usepackage{graphicx}
\usepackage{bm}
\usepackage{siunitx}
\usepackage{amsmath}
\usepackage{multirow}
\usepackage[FIGTOPCAP]{subfigure}
\usepackage{epstopdf}	       % Included EPS files automatically converted to PDF to include with pdflatex

\DeclareSIUnit{\sqrthz}{\ensuremath{\sqrt{\text{\hertz}}}}
\DeclareSIUnit{\voltnoise}{\volt\per\sqrthz}

\newcommand\T{\rule{0pt}{2.6ex}}       % Top strut
\newcommand\B{\rule[-1.2ex]{0pt}{0pt}} %Bottom strut

\begin{document}

\title{Contactless and absolute linear displacement detection based upon 3D printed magnets combined with passive radio-frequency identification}% Force line breaks with \\
%\thanks{Something?}%

\author{Roman Windl}
\email{roman.windl@univie.ac.at}
\author{Claas Abert}
\author{Florian Bruckner}
\author{Christian Huber}
\affiliation{CD-Laboratory: Advanced Magnetic Sensing and Materials, University of Vienna, Physics of Functional Materials, W\"ahringer Stra\ss e 17, 1090 Vienna, Austria  }%Lines break automatically or can be forced with \\
\author{Christoph Vogler}
\affiliation{University of Vienna, Physics of Functional Materials, W\"ahringer Stra\ss e 17, 1090 Vienna, Austria }
\author{Herbert Weitensfelder}
\author{Dieter Suess}
  
 \affiliation{CD-Laboratory: Advanced Magnetic Sensing and Materials, University of Vienna, Physics of Functional Materials, W\"ahringer Stra\ss e 17, 1090 Vienna, Austria    }%Lines break automatically or can be forced with \\


\begin{abstract}
Within this work a passive and wireless magnetic sensor, to monitor linear displacements is proposed. We exploit recent advances in 3D printing and fabricate a polymer bonded magnet with a spatially linear magnetic field component corresponding to the length of the magnet. Regulating the magnetic compound fraction during printing allows specific shaping of the magnetic field distribution. A giant magnetoresistance magnetic field sensor is combined with a radio-frequency identification tag in order to passively monitor the exerted magnetic field of the printed magnet. 
Due to the tailored magnetic field, a displacement of the magnet with respect to the sensor can be detected within the sub-\si{\milli\metre} regime.
The sensor design provides good flexibility by controlling the 3D printing process according to application needs. Absolute displacement detection using low cost components and providing passive operation, long term stability and longevity renders the proposed sensor system ideal for structural health monitoring applications.
\end{abstract}

%\keywords{passive; RFID; sensor; position; }

\maketitle

\begin{figure}[h]
\centering
\includegraphics[width=\columnwidth]{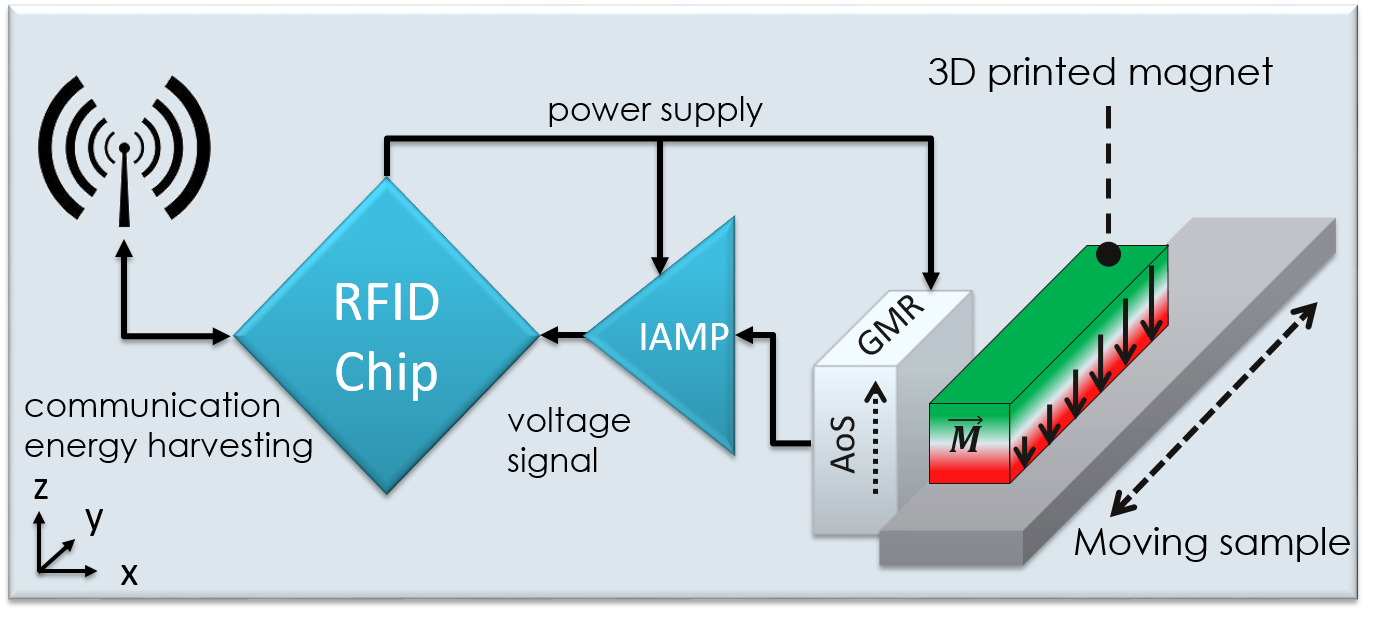}% Here is how to import EPS art
\caption{\label{fig:setup} A schematic of the sensor system is illustrated. The magnetization of the 3D printed magnet points along negativ z-axis. A GMR sensor with an axis of sensitivity (AoS) along positive z-axis observes the resulting spatially linear magnetic field change if the sample is moved along y-axis.}
\end{figure}

%\section{\label{sec:Int}Introduction}
Linear displacement systems in the \si{\milli\metre} regime are widely used among different industries. A huge variety of measuring techniques is currently available, for example infra-red, ultra-sonic, magnetic~\cite{disp_mag,disp_GMR,disp_ha}, optic~\cite{disp_optic} or even digital picture processing. Applications which demand passive, wireless and long term operation, like for example structural health monitoring, are the main scope of the presented displacement detection approach. 
The proposed sensor setup consists of three main parts: (i) an radio-frequency identification (RFID) tag, (ii) a giant magnetoresistance (GMR) magnetic field sensor combined with an instrumentation amplifier (IAMP) and (iii) a 3D printed polymer bonded magnet, shown in FIG.~\ref{fig:setup}. With a combination of (i), and (ii) it is possible to passively monitor different physical properties like temperature~\cite{Meins1} or strain~\cite{Meins2}.

This work presents a displacement detection technique based on 3D printed magnets with a spatially linear magnetic field component at a specific region along the stretched axis of the magnet. The approach of varying the magnetic compound fraction $\rho_m$ by 3D printing polymer bonded magnetic materials allows magnetic field shaping as presented by Huber et al.~\cite{Huber2}. Designing the magnetic field magnitudes according to the characteristics of the used magnetic field sensor increases measurement resolution and enables new types of applications. The main advantages of the presented sensor method are the absolute displacement detection, passive operation, commercially available and low cost components, long term stability, longevity and low soil influence. Hence, hazardous or very dusty environments are possible areas of application.

A 3D printer with a mixing extruder is used in order to fabricate a polymer bonded magnet with varying magnetic compound fraction. Supplying the mixing extruder with the nylon PA6 and a hard ferrite ($SrOx_6Fe_2O_3$) inside a PA6 matrix called Sprox \textsuperscript{\textregistered} 10/20p from Magnetfabrik Bonn. Since the magnetic field range of the GMR sensor is in the \si{\milli\tesla} regime, a hard ferrite is a suitable permanent magnet.
Parameters like the start compound fraction $\rho_{m,\mathrm{0}}$, stop compound fraction $\rho_{m,\mathrm{l_y}}$, and magnetization are tuned to match a specific magnetic field magnitude at these points. The magnetic compound fraction is varied from $\rho_{m,\mathrm{0}} = \SI{10}{\percent}$, resulting in composition of \SI{10}{\percent} Sprox \textsuperscript{\textregistered} 10/20p and \SI{90}{\percent} PA6, to $\rho_{m,\mathrm{l_y}} = 80\%$ in order to match the linear range of the used GMR sensor. Due to printing irregularities the true size of the magnet, after the printing process, is \SI{9.5x39x9.5}{\milli\metre} ($l_x \times l_y \times l_z$) instead of \SI{10x40x10}{\milli\metre} ($l_x \times l_y \times l_z$). An electromagnet is used to magnetize the 3D printed cuboid with $B_z = \SI{-0.3}{\tesla}$. $y$ describes the position along the y-axis of the magnet. 

\begin{figure}[!h]
	\centering
	\includegraphics[width=\columnwidth]{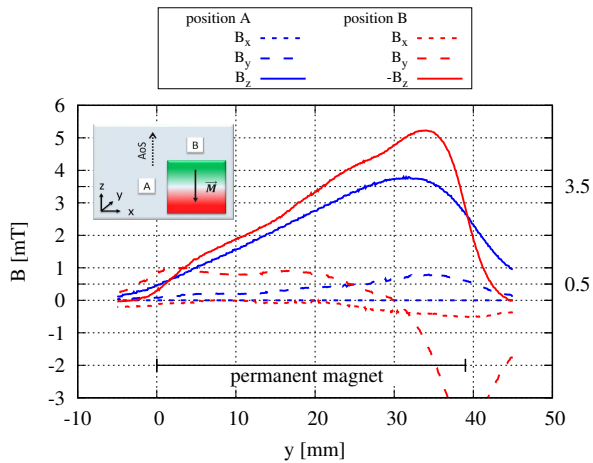}% Here is how to import EPS art
	\caption{\label{fig:mag_line_scan} The magnetic field is measured along the y-axis with a 3D hall probe at a distance of \SI{2.2}{\milli\metre}. Because of printing irregularities the $-B_z$ component at position B (solid red line) does not increase perfectly linear.}		
\end{figure}

Huber et al.~\cite{Huber1} showed how to upgrade the 3D printer with a TLV493D 3D magnetic field sensor from Infinenon Technologies to perform a 3D scan of the magnetic field produced by an arbitrary sample. The same method is used to characterize the linearity of the magnetic field induced by the magnetization of the 3D printed magnet, and is illustrated in FIG.~\ref{fig:mag_line_scan}. 

Positioning the GMR sensor is very crucial in order to measure the spatially linear magnetic field component $B_z$. GMR sensors are in plane sensitive and placing the sensor at position A, see inset Fig.~\ref{fig:mag_line_scan}, allows larger distances between the permanent magnet and the sensor. 
Additionally, this position is more suitable due to the significantly smoother $B_z$ component, and smaller influences of the $B_x$ and $B_y$ components. The solid red line represents $-B_z$ at position B above the magnet from $y = \SI{5}{\milli\metre}$ to $y = \SI{33}{\milli\metre}$ where a nearly linear increase is measurable. A noticeable deviation of linearity is recognized from $y = \SI{15}{\milli\metre}$ to $y = \SI{28}{\milli\metre}$ as a result of printing irregularities. However, at position A, sideways along the magnet, the $B_z$ component represented by a solid blue line matches the GMR sensors linear range characteristics from $y = \SI{0}{\milli\metre}$ with $B_z = \SI{0.5}{\milli\tesla}$ to $y = \SI{26}{\milli\metre}$ with $B_z = \SI{3.5}{\milli\tesla}$.

Within the following paragraphs a detailed description of the single sensor components is provided. As RFID tag the SL900A~\cite{SL900A} from AMS AG is used. The external sensor front-end (SFE) allows different voltage adjustments and ranges to offer a large variety of applications for the two analog inputs. 

\begin{figure}[!h]
	\centering
	\includegraphics[width=\columnwidth]{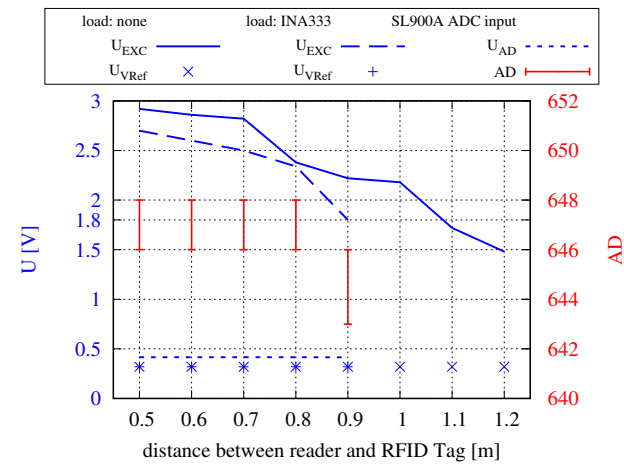}% Here is how to import EPS art
	\caption{\label{fig:range} Output voltage of the EXC pin $U_{\mathrm{EXC}}$ (solid and long dashed blue line) over the distance between the Astra-EX reader antenna and the SL900A. The $V_{REV}$ pin voltag  $U_{\mathrm{VRef}}$ (triangle up and triangle down in blue color) is stable until no communication is possible at all. Blue color indicates voltages and the red bars indicate the $RFID_{\mathrm{EXT1}}$ range. The IAMP output voltage (blue short dashed line) is steady over the whole voltage supply range. }	
\end{figure}

Because of the RFID tags analog to digital conversion with 10 bit the maximum value is $AD_{\mathrm{max}}=2^{10}=1024$. Energy harvested by the reader field is converted in order to supply external sensors by the output supply voltage $U_{\mathrm{EXC}}$. Because of the low current supplied $I_{\mathrm{EXC}} < \SI{200}{\micro\ampere} $ most Hall sensors are not applicable. Hence, GMR sensors with resistances $R > \SI{10}{\kilo\ohm}$ are used to keep the power consumption as low as possible. 
As a magnetic field sensor the GMR AA006 from NVE Corporation with the following characteristics is used: $R_{\mathrm{GMR}} = \SI{30}{\kilo\ohm} \pm \SI{20}{\percent}$, a linear range from $\SI{0.5}{\milli\tesla} < B_{\mathrm{ext}} < \SI{3.5}{\milli\tesla}$, and saturation at $B_{\mathrm{sat}}\approx\SI{5}{\milli\tesla}$. A downside of the used GMR sensor is a hysteresis of approximately \SI{4}{\percent} at unipolar operation.
Due to the fact that GMR sensors are configured in a Wheatstone bridge setup two signal inputs are required. 
The measurable voltage difference is too small for the SFE characteristics and therefore, an instrumentation amplifier helps to overcome this issue. The INA333 from Texas Instruments is an appropriate choice because of the low power consumption and the wide supply voltage. An additional advantage of using an instrumentation amplifier is that the GMR sensors output voltage can be adjusted to nearly match the specific analog input characteristics of the RFID tag. Supplying the RFID tag and other sensor components through the reader field decreases the reading distance due to the increased power consumption, as shown in Fig.~\ref{fig:range}. 
As an RFID reading device the Astra-EX from ThingMagic is used. If the distance to the reader is increased a decreasing output supply voltage $U_{\mathrm{EXC}}$ of the RFID tag is monitored. Therefore, the GMR sensor is supplied by the $U_{\mathrm{VRef}}$ voltage, instead of the distance dependent EXC pin voltage $U_{\mathrm{EXC}}$. Additionally, reducing the power consumption due to the lower GMR sensors supply voltage $U_{\mathrm{VRef}}$. When the EXC pin voltage reaches \SI{1.8}{\volt} the $AS$ uncertainty increases and the mean is decreased. After $U_{\mathrm{EXC}}$ drops below the minimum IAMP supply voltage of \SI{1.8}{\volt}, communication with the RFID tag is nearly impossible. Keeping the power consumption as low as possible is crucial to allow passive operation. Therefore, perfectly adjusting the GMR sensors output voltage by further electronic components is not considered. 

In order to measure the real magnetic field dependence and resolution, the presented sensor system is calibrated inside a Helmholtz Coil. The distance between the RFID reader Astra-EX and the SL900A tag is \SI{0.5}{\metre} for all following measurements.

\begin{figure}[!h]
	\centering
	\includegraphics[width=\columnwidth]{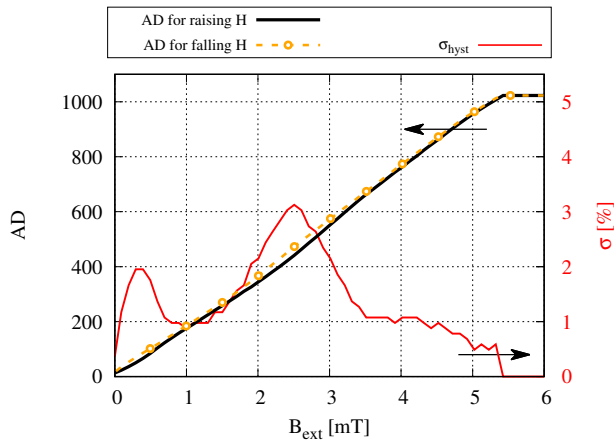}% Here is how to import EPS art
	\caption{\label{fig:calib} The transfer function for the proposed sensor setup is shown. For raising H (black solid line, left axis) and falling H (orange dashed pointed line, left axis) a hysteresis induced measurement error $\delta_{\mathrm{hyst}}$ (solid red line, right axis) is measured.   }	
\end{figure}

The external magnetic field $B_{\mathrm{ext}}$ is altered from \SIrange{6}{0}{\milli\tesla} and backwards with \SI{0.1}{\milli\tesla} steps. Fig.~\ref{fig:calib} shows the measurement results and reveals hysteric behaviour.

The performed measurements at room temperature indicate a maximum measured magnetic field $B_{\mathrm{ext,max}} = \SI{5.4}{\milli\tesla}$ for the maximum analog to digital value $AD_{\mathrm{max}} = 1023$. If the transfer function from Fig.~\ref{fig:calib} is fitted by a polynomial equation $T_{\mathrm{B}}$, recalculation of the external applied magnetic field 
\begin{equation}
\label{eq:T_B}
B_{\mathrm{ext}} = T_{\mathrm{B}}(AD)
\end{equation}
is possible. In order to estimate the correlation between the y-axis position $y_{\mathrm{pos}}$ of the 3D printed magnet, and the magnetic field value $B$ at this position, a second transfer function $T_{\mathrm{M}}$ is introduced
\begin{equation}
y_{\mathrm{pos}} = T_{\mathrm{M}}(B) = \frac{\Delta l}{\Delta B_\mathrm{l}}B.
\label{eq:T_M}
\end{equation}
By combining equation~\ref{eq:T_B}, and~\ref{eq:T_M}, the conversion of $AD$ into a y-axis position
\begin{equation}
y_{\mathrm{pos}} = T_{\mathrm{M}}(T_{\mathrm{B}}(AD))
\label{eq:y_pos}
\end{equation}
is achieved, allowing the recalculation of the magnets y-axis displacement with respect to the sensor position.

\begin{figure}[htb]
	\centering
	\includegraphics[width=\columnwidth]{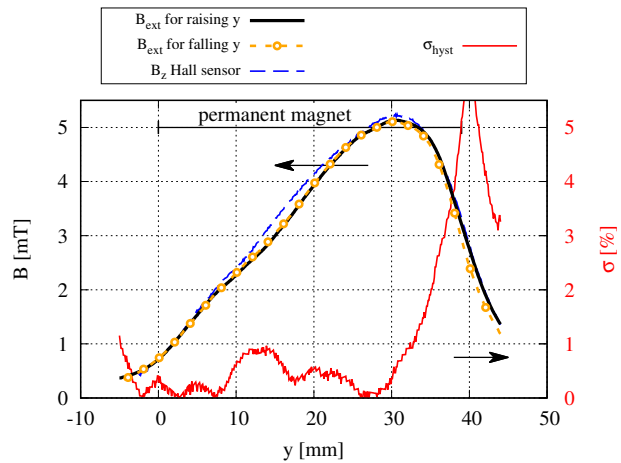}% Here is how to import EPS art
	\caption{\label{fig:result_mT} $B_z$ for the Hall Sensor TLV493D (blue dashed line, left axis) is compared to the recalculated $B_{\mathrm{ext}}$ for raising y from \SIrange{-5}{44}{\milli\metre} (black solid line, left axis) and falling y from \SIrange{44}{-5}{\milli\metre} (orange dashed pointed line, left axis). $\delta_{\mathrm{hyst}}$ (red solid line, right axis) illustrates the hysteric influence. }	
\end{figure}

The distance between the GMR sensor and the permanent magnet regulates the maximum magnetic field, and therefore is used to fit the GMR sensor range.  
Because the sensor calibration showed linearity up to \SI{5.4}{\milli\tesla} the distance between the GMR sensor and the permanent magnet is decreased to \SI{1.8}{\milli\metre}, in order to utilize the whole linear range of the GMR sensor. With the GMR sensor mounted upon the 3D printers head, magnetic field measurements along the y-axis of the magnet are performed. By using the transfer function $ T_{\mathrm{B}}$ from equation~\ref{eq:T_B}, recalculation of the external magnetic field $B_{\mathrm{ext}}$ is performed and illustrated in Fig.~\ref{fig:result_mT}.

The AD resolution is given by  
\begin{equation}
\label{eq:s_res}
\sigma_{\mathrm{res}} = \frac{1}{AD_{\mathrm{max}}} = \SI{0.097}{\percent}.
\end{equation}

$\Delta AD_{\mathrm{total}}$ consists of all possible noise sources noted by
\begin{equation}
\begin{split}
\label{eq:ad_tot}
\Delta AD_{\mathrm{total}}^2 &= \Delta AD_{\mathrm{ADC}}^2+\Delta AD_{\mathrm{IAMP}}^2\\
&+\Delta AD_{\mathrm{GMR}}^2+\Delta AD_{\mathrm{ext}}^2 
\end{split}
\end{equation}
where the sources are represented by $\Delta AD_{\mathrm{ADC}}$  the analog to digital converter noise, $\Delta AD_{\mathrm{IAMP}}$ the noise from the IAMP, the GMR sensor noise $\Delta AD_{\mathrm{GMR}}$, and external noise $\Delta AD_{\mathrm{ext}}$. A detectivity of $\approx \SI{10}{\nano\tesla\per\sqrthz}$ at $f=\SI{0.1}{\hertz}$ was measured by Stutzke et al.~\cite{detGMR} for the AA002 which has an approximately three times higher mean sensitivity than the AA006 rendering $\Delta_{\mathrm{GMR}}$ negligible. The noise produced by the instrumentation amplifier, $\approx \SI{3}{\micro\volt}$ for DC operation with Gain $G_{\mathrm{IAMP}} = 1$, is insignificant because it is clearly smaller than the analog to digital resolution. Hence, only  $\Delta AD_{\mathrm{ADC}}$ and $\Delta AD_{\mathrm{ext}}$ are the main noise sources for the proposed measurement setup. 
For each measurement step hundred $AD$ measurements are averaged, deviating by a noise induced minimum measurement error 
\begin{equation}
\sigma_{\mathrm{noise}} = \sqrt{\frac{\Delta AD_{\mathrm{tot}}^2}{AD_{\mathrm{max}}^2}} =  \SI{0.293}{\percent}
\label{eq:s_noise}
\end{equation}

Fig.~\ref{fig:result_mT} indicates nearly linear behaviour in region $l_1$ from $l_{1,\mathrm{min}} = \SI{0}~$to$~l_{1,\mathrm{max}} =  \SI{30}{\milli\metre}$. Therefore, $l_1$ is used as displacement detection region.
Within this region the hysteresis induced measurement error
\begin{equation}
\sigma_{\mathrm{hyst}} = \frac{AD_{\mathrm{hyst}}}{AD_{\mathrm{max}}} = \SI{0.966}{\percent}
\label{eq:s_hyst}
\end{equation}
is computed. The sensor characteristic properties are set in relation to displacement detection region $l_1$ within TABLE~\ref{tab:d_l-calc}.
\begin{table}[h]
	\caption{\label{tab:d_l-calc} Calculation of length relations for $l_{1,\mathrm{min}} = \SI{0}~$, and $~l_{1,\mathrm{max}} =  \SI{30}{\milli\metre} $.}
		\begin{ruledtabular}
		\begin{tabular}{ c | c c c}
			\multirow{2}{*}{type} & $\sigma$ & $\Delta l$ \\ 	
			& [\si{\percent}] & [\si{\micro\meter}] &\T\B \\ \hline

			res & 0.097 &  29.1 & \T \\ 
			noise & 0.293 &  87.9 & \\ 
			hyst & 0.966 &  289.8 & \\ 
						
		\end{tabular} 
	\end{ruledtabular}
\end{table}

Therefore, the detection of displacements within the sub-\si{\milli\metre} regime is possible. The magnet can be tuned towards application needs, for example it can be elongated or shortened to achieve predefined length resolutions with the advantage of absolute positioning. Additionally, the magnetic field can be shaped in order to reduce or even compensate the non linearity of the used GMR sensor by a specific compound fraction $\rho_m$ variation. Hence, disadvantages of magnetic field sensor can be incorporated by system design, reducing post calculation complexity and therefore increasing the systems reliability. 
A hysteresis free GMR magnetic field sensor $\sigma_{\mathrm{hyst}}  = 0$ as presented by Brueckl et al.~\cite{Vortex1} and described by a patent~\cite{Vortex2} significantly decreases the detectable displacements.\\

In the following, we summarize the proposed displacement detection system. It consists of three main components (i) an RFID tag, (ii) a GMR magnetic field sensor combined with an IAMP, and (iii) a 3D printed polymer bonded magnet with linear magnetic compound fraction variation along the y-axis. The digital value $AD$ of the analog to digital input of the RFID tag correlates to the magnets $B_z$ component at a specific y-axis position. Main characteristics of the presented system are a resolution $\sigma_{\mathrm{res}}$ of $\SI{0.097}{\percent}$, a measurement uncertainty $\sigma_{\mathrm{noise}}$ of $\SI{0.293}{\percent}$, and a hysteresis induced measurement uncertainty $\sigma_{\mathrm{hyst}}$ of $\SI{0.966}{\percent}$. Hence, displacements within sub-\si{\milli\metre} regime are detectable. The magnet can be tuned towards magnetic sensor characteristics in order to utilize the whole linear range. Additionally, further adjustments of the magnet towards application needs are possible, for example decreasing or increasing the length of the overall position detection. The precision of the presented system can be improved by replacing the used GMR sensor by a hysteresis free GMR sensor~\cite{Vortex1,Vortex2}, rendering $\sigma_{\mathrm{hyst}} = 0$. The sensor design provides good flexibility adjusting to application needs by controlling the 3D printing process. Structural health monitoring is the main scope of application for the proposed passive, low cost, long term stable, and absolute displacement detection system.

%\begin{acknowledgments}
The authors would like to gratefully acknowledge the Christian Doppler Laboratory Advanced Magnetic Sensing and Materials, for financial support. The Laboratory is financed by the Austrian Federal Ministry of Science, Research, and Economy, and the  National Foundation for Research, Technology, and Development.
%\end{acknowledgments}

\appendix

\bibliography{paper_windl}% Produces the bibliography via BibTeX.

\end{document}